\documentstyle[12pt,epsf,graphics]{article}
\begin{document}
\title{Shortcuts for Graviton Propagation in a Six Dimensional Brane
World Model}
\author{Elcio Abdalla, Adenauer Casali and Bertha Cuadros-Melgar \\ \\
Instituto de F\'\i sica, Universidade de S\~ao Paulo \\
  C.P.66.318, CEP 05315-970, S\~ao Paulo, Brazil}
\date{}
\maketitle
\begin{abstract}
 We consider a six dimensional brane world model with asymmetric warp
 factors for time and both extra spatial coordinates, $y$ and
 $z$. We derive the set of differential equations governing the
 shortest graviton path and numerically solve it for AdS-Schwarzschild
 and AdS-Reissner-Nordstr\"om bulks. In both cases we derive a
 set of conditions for the existence of shortcuts in bulks with
 shielded singularities and show some examples of shortcuts obtained
 under these conditions. Consequences are discussed. 
\end{abstract}



\newpage

\section{Introduction} 

The ideas of Kaluza and Klein \cite{k-k}, advocating the physical
possibility of extra dimensions in order to achieve the unification of
different field theories can be considered as a landmark in Quantum
Field Theory. 

Such an importance grew specially half a century after the original works,
in the framework of supergravity and string theories. In the latter,
the existence of extra dimensions is actually enforced by consistency.

Furthermore, new possibilities to realize the extra dimensions permitted
to explore new mechanisms of explaining unified field theories. The
possibility of explaining hierarchies in such a context is specially
appealing and has been confirmed in the works of Arkani-Hamed and
others \cite{add,antoniadis}. The hierarchy between the electro-weak
($\sim 100 \,GeV$) and the Planck ($10^{19} \,GeV$) scales has been
focused by means of the consideration of extra dimensions at a
submilimeter size, which shows up in a theory of two extra dimensions
connecting both scales. Such an idea replaces the usual one where
extra dimensions should only show up at the Planck scale, and the tower
of massive particles thus generated is above that level and has a
wider validity, including cosmology \cite{acb-kl}.

Such size constraints on the size of the extra dimensions constitute
drawbacks in the formulation of the theory.

More recently, Randall and Sundrum \cite{rs1,rs2} proposed a model -or
rather a class of models- where there is a warp factor in the metric,
such that even infinitely large extra dimensions are allowed.

The existence of large extra dimensions with a warp factor naturally
raises the question of whether information can follow a shorter path
outside the brane riding on gravitons
\cite{ishihara,cald-lang,freese,chkorio}. 
We proposed a simple calculation to establish the shortest
path followed by a graviton \cite{acmfw}, which propagating in all
dimensions in the so called bulk, could in principle follow a path
which decouples from the brane, that is, from our universe, returning
later to another point, advanced in time with respect to a photon, which by
construction must follow a path in the brane, thus being delayed. In
our previous paper we considered a model constructed in
\cite{binetruy2} which was basically a generalized
Friedmann-Robertson-Walker universe with cosmological constant, with
different scenarios in the brane (where are living the Standard
Model fields) and in the bulk. The result was actually a negative one,
that is, the shortest path followed by the graviton was the same as the one
followed by the photon, namely, inside the brane.

In a model introduced by Cs\'aki et. al. \cite{csaki1} the speed of
light along flat four dimensional sections varies over the extra
dimensions due to different warp factors for the space and time
coordinates, a construction similar to the one of Randall and
Sundrum. Thus the authors proposed that gravitational waves might
travel faster than photons, which remain in the brane. The delay between
electromagnetic and gravity waves may be experimentally detected with
the gravitational waves detectors under way \cite{chkorio,clayton}. 

The models are basically AdS-Schwarzschild or AdS-Reissner-Nordstr\"om black
holes in the bulk. Brane models in AdS space with Schwarzschild
singularities have been used to understand the AdS/CFT correspondence
and looks like a promising theoretical model \cite{aws}. They are based on
the Randall-Sundrum scheme \cite{rs1,rs2}, 
where a large mass hierarchy is obtained with uncompactified 
dimensions from solutions of Einstein equations in higher dimensions
(i.e., in the bulk) with two separated branes. The four dimensional
part of the metric is multiplied by a ``warp'' factor which is a
rapidly changing function of the additional dimension.

In this paper we consider a six-dimensional model and look for possible
shortcuts for AdS-Schwarzschild and AdS-Reissner-Nordstr\"om bulk
configurations. The paper is organized as follows. In section 2 we
describe a general six-dimensional model, derive Einstein equations,
and find the Israel conditions the metric has to satisfy due to the
brane embedding. At this point we choose a metric describing a
six-dimensional black hole and add a $Z_2$ symmetry. In
section 3 we find the Euler-Lagrange equations which define the
graviton path in this model. Section 4 is devoted to study the
numerical solutions of these equations in the context of
AdS-Schwarzschild bulk finding certain analytical requirements for the
existence of shortcuts. The AdS-Reissner-Nordstr\"om bulk is studied 
in section 5, where we perform an analytical discussion to impose a
set of conditions under which shortcuts can coexist with shielded
singularities. Finally, consequences are discussed in section 6. 
 
\section{A Six-Dimensional Model}

We consider a six-dimensional model, such as the one constructed
by Kanti et. al. \cite{kanti}. We also search for a solution of
six-dimensional Einstein equation in AdS space of the form

\begin{equation}\label{metric}
ds^2 = -n^2(t,y,z) dt^2 + a^2(t,y,z) d\Sigma_{k} ^2 + b^2(t,y,z)
\left\{ dy^2 + c^2(t,y,z) dz^2 \right\}
\end{equation}
where $d\Sigma_{k} ^2$ represents the metric of the three dimensional
spatial sections with $k=-1,\,0,\,1$ corresponding to a hyperbolic, a
flat and an elliptic space, respectively.

The components of the Einstein tensor read

\begin{eqnarray}\label{g00}
G_{00}&=& \scriptstyle{3 { \dot a \over a} \left( {\dot a \over a} +
    2{\dot b \over 
    b} + {\dot c \over c} \right) + {\dot b \over b} \left({\dot b
    \over b} + {\dot c \over c} \right) - {n^2 \over b^2} \left\{
    \left( 3 {{\partial_y ^2 a} \over a} + 3 {{(\partial_y a)^2}\over
    a^2} - {{(\partial_y b)^2} \over b^2} + {{\partial_y ^2 b} \over
    b} + {{\partial_y ^2 c} \over c} + 3 {{\partial_y a \partial_y
    c}\over{ac}} + {{\partial_y b \partial_y c}\over{bc}} \right) +
    \right.} \nonumber\\
&& \scriptstyle{\left.+{1
    \over c^2} \left( 3 {{\partial_z ^2 a} \over a} +3 {{(\partial_z a)^2}\over
    a^2} - {{(\partial_z b)^2} \over b^2} + {{\partial_z ^2 b} \over
    b} - 3 {{\partial_z a \partial_z c}\over{ac}} - {{\partial_z b
    \partial_z c}\over{bc}} \right) \right\} }
\end{eqnarray}
\begin{eqnarray}\label{gij}
G_{ij}&=& \scriptstyle{ {a^2 \over n^2} \left\{ -2 {\ddot a \over a}
    -2{\ddot b \over b}  - {\ddot c \over c} - {{\dot a^2} \over a^2}
    - {{\dot b^2} \over b^2} + 2 {{\dot a \dot n}\over{an}} - 4 {{\dot a
        \dot b}\over{ab}} - 2 {{\dot a \dot c}\over{ac}} + 2 {{\dot b
        \dot n}\over{bn}} + {{\dot c \dot n}\over{cn}} -3 {{\dot b
        \dot c}\over{bc}}\right\} \delta_{ij} + }\nonumber \\
&& \scriptstyle{ + {a^2 \over b^2} \left\{ \left( 2 {{\partial_y ^2
          a}\over a} + {{(\partial_y a)^2}\over a^2} + {{\partial_y ^2
          b}\over b} - {{(\partial_y b)^2}\over b^2} + {{\partial_y ^2
          n}\over n} + {{\partial_y ^2 c}\over c} + 2 {{\partial_y a
          \partial_y n}\over{an}} + 2 {{\partial_y a \partial_y
          c}\over{ac}} + {{\partial_y c \partial_y n}\over{cn}} +
      {{\partial_y b \partial_y c}\over{bc}} \right) + \right. \,\,\,\,}
\nonumber \\ 
&& \scriptstyle{ \left. + {1 \over c^2} \left( 2 {{\partial_z ^2
          a}\over a} + {{(\partial_z a)^2}\over a^2} + {{\partial_z ^2
          b}\over b} - {{(\partial_z b)^2}\over b^2} + {{\partial_z ^2
          n}\over n} + 2 {{\partial_z a \partial_z n}\over{an}} - 2
      {{\partial_z a \partial_z c}\over{ac}} - {{\partial_z c
          \partial_z n}\over{cn}} - {{\partial_z b \partial_z
          c}\over{bc}} \right) \right\} \delta_{ij} }
\end{eqnarray}
\begin{eqnarray}\label{g55}
G_{55}&=& \scriptstyle{ -{b^2 \over n^2} \left( {\ddot b \over b} + 3
    {\ddot a \over a} + {\ddot c \over c} - {{\dot b \dot n}\over{bn}}
    + 3 {{\dot a \dot b}\over{ab}} + 2 {{\dot b \dot c}\over{bc}} - 3
    {{\dot a \dot n}\over{an}} - {{\dot c \dot n}\over{cn}} + 3 {{\dot
    a \dot c}\over{ac}} + 3 {{\dot a^2} \over a^2} \right) + }
    \nonumber \\
&& \scriptstyle{ + 3 {{(\partial_y a)^2}\over a^2} + 3 {{\partial_y a
    \partial_y b}\over{ab}} + 3 {{\partial_y a \partial_y c}\over{ac}}
    + 3 {{\partial_y a \partial_y n}\over{an}} + {{\partial_y b
    \partial_y n}\over{bn}} + {{\partial_y c \partial_y n}\over{cn}} +
    \qquad \qquad \qquad \qquad \quad \quad} \nonumber \\
&& \scriptstyle{ + {1 \over c^2} \left\{ 3 {{\partial_z ^2 a}\over a} +
    3 {{(\partial_z a)^2}\over a^2} + {{\partial_z ^2 n}\over n} -
    \left( 3 {{\partial_z a \partial_z b}\over{ab}} + 3 {{\partial_z a
    \partial_z c}\over{ac}} - 3 {{\partial_z a \partial_z n}\over{an}}
    + {{\partial_z b \partial_z n}\over{bn}} + {{\partial_z c
    \partial_z n}\over{cn}} \right) \right\}\, } 
\end{eqnarray}
\begin{eqnarray}\label{g66}
G_{66}&=& \scriptstyle{ -{{b^2 c^2}\over n^2} \left( 3 {\ddot a \over
      a} + {\ddot b \over b} - {{\dot b \dot n}\over {bn}} + 3 {{\dot
      a \dot b}\over {ab}} - 3 {{\dot a \dot n}\over {an}} + 3 {{\dot
      a^2}\over a^2} \right) + } \nonumber \\
&& \scriptstyle{ + c^2 \left\{ 3 {{\partial_y ^2 a}\over a} + 
      {{\partial_y ^2 n}\over n} + 3 {{(\partial_y a)^2}\over a^2} + 3
      {{\partial_y a \partial_y n}\over{an}} - 3 {{\partial_y a
      \partial_y b}\over{ab}} - {{\partial_y b \partial_y n}\over{bn}}
      \right\} + \qquad \qquad \qquad \qquad \qquad } \nonumber \\
&& \scriptstyle{ + 3 {{(\partial_z a)^2}\over a^2} + 3 {{\partial_z a
      \partial_z b}\over{ab}} + 3 {{\partial_z a \partial_z
      n}\over{an}} + {{\partial_z b \partial_z n}\over{bn}} }
\end{eqnarray}
\begin{eqnarray}\label{g05}
G_{05}&=& \scriptstyle{ -3 {{\partial_y \dot a}\over a} - {{\partial_y
      \dot b}\over b} -  {{\partial_y \dot c}\over c} + 3 {\dot a
      \over a} \left({{\partial_y n}\over n} \right) + {\dot b \over
      b} \left( {{\partial_y n}\over n} + 3 {{\partial_y a}\over a} +
      {{\partial_y b}\over b} \right) + {\dot c \over c} \left(
      {{\partial_y n}\over n} - {{\partial_y b}\over b} \right) \qquad
      \qquad \quad \quad \,\,\,\,\,\,\,}
\end{eqnarray}
\begin{eqnarray}\label{g06}
G_{06}&=& \scriptstyle{ -3 {{\partial_z \dot a}\over a} - {{\partial_z
      \dot b}\over b} + 3 {\dot a
      \over a} \left({{\partial_z n}\over n} \right) + {\dot b \over
      b} \left( {{\partial_z n}\over n} + 3 {{\partial_z a}\over a} +
      {{\partial_z b}\over b} \right) + {\dot c \over c} \left( 3
      {{\partial_z a}\over a} + {{\partial_z b}\over b} \right) \qquad
      \qquad \qquad \qquad \quad \,\, }
\end{eqnarray}
\begin{eqnarray}\label{g56}
G_{56}&=& \scriptstyle{ - {{\partial_z \partial_y n}\over n} -3
  {{\partial_z \partial_y a}\over a} + {{\partial_z b \partial_y
  n}\over{bn}} + 3 {{\partial_y a \partial_z b}\over{ab}} +
  {{\partial_y b \partial_z n}\over{bn}} + {{\partial_y c \partial_z
  n}\over{cn}} + 3 {{\partial_z a \partial_y b}\over{ab}} + 3
  {{\partial_z a \partial_y c}\over{ac}} \qquad \qquad \quad \,\,\,\,\,} 
\end{eqnarray}

The total energy-momentum tensor can be decomposed in two parts
corresponding to the bulk and the brane as
\begin{equation}\label{emtensor}
\tilde T^M _N = \breve T^{M(B)} _N + T^{M (b)} _N \, ,
\end{equation}
where the brane contribution can be written as
\begin{equation}\label{brane}
T^{M (b)} _N = { {\delta (z-z_0)} \over {bc}} \,diag \,(-\rho,
p,p,p,\hat p, 0) \, .
\end{equation}

In order to have a well-defined geometry, the metric must be
continuous across the brane; however, its derivatives with respect to
$z$ can be discontinuous at the position of the brane, generating a
Dirac $\delta$-function in the second derivatives of the metric with
respect to $z$ \cite{binetruy1}. These $\delta$ function terms must be
matched with the components of the brane energy-momentum tensor
(\ref{brane}) in order to satisfy Einstein equations. Thus, using
(\ref{g00}), (\ref{gij}) and (\ref{g55}) we obtain the following
Israel conditions, 
\begin{eqnarray}\label{israel}
{{[\partial_z a]} \over {a_0 b_0 c_0}} &=& -{{\kappa_{(6)} ^2} \over 4}
(p-\hat p + \rho) \, ,\nonumber \\
{{[\partial_z b]} \over {b_0 ^2 c_0}} &=& -{{\kappa_{(6)} ^2} \over 4}
\left\{ \rho - 3(p-\hat p) \right\} \, ,\\
{{[\partial_z n]} \over {b_0 c_0 n_0}} &=& {{\kappa_{(6)} ^2} \over 4}
\left\{ \hat p + 3 (p+ \rho) \right\} \, . \nonumber
\end{eqnarray}

A metric of the form (\ref{metric}) satisfying six dimensional Einstein
equations is given by
\begin{equation}\label{bhmetric}
ds^2 = - h(z) dt^2 + {z^2 \over l^2} d\Sigma_k ^2 + h^{-1}
(z) dz^2 \, ,
\end{equation}
where
\begin{equation}\label{spacediff}
d\Sigma_k ^2 = {{dr^2} \over {1-kr^2}} + r^2 d\Omega_{(2)} ^2 +
(1-kr^2) dy^2 \, ,
\end{equation}
and
\begin{eqnarray}
&h(z) = k + {z^2 \over l^2} - {M \over z^3}& \, ,\quad  \hbox{for
AdS-Schwarzschild bulk} \, ,\label{hsch} \\
&h(z) = k + {z^2 \over l^2} - {M \over z^3} + {Q^2 \over z^6}& \, ,\quad
\hbox{for AdS-Reissner-Nordstr\"om bulk} \, , \label{hrn} 
\end{eqnarray}
with $l^{-2} \propto -\Lambda$ ($\Lambda$ being the cosmological
constant), which describes a black hole in the bulk, located at $z=0$.  

Following \cite{csaki1}, we find a further solution by means of a
$Z_2$ symmetry inverting the space with respect to the brane
position. That is, considering a metric of the form 
\begin{equation}\label{genmetric}
ds^2 = -A^2(z) dt^2 + B^2(z) d\Sigma_{(4)} ^2 + C^2(z) dz^2
\end{equation}
and the brane to be defined at $z=z_0$, there is a solution given by
\begin{eqnarray}\label{z2sym}
&A(z), \,\, B(z), \,\, C(z)& \, ,\quad \hbox{for} \quad z\leq z_0 \,
,\nonumber\\ 
&A(z_0 ^2/z), \,B(z_0 ^2/z), \,C(z_0 ^2/z) {{z_0 ^2} \over z^2}& \, ,\quad
\hbox{for} \quad z \geq z_0 \, .
\end{eqnarray} 
The $Z_2$-symmetry corresponds to $z \rightarrow z_0 ^2/z$.

The static brane still has to obey the Israel conditions
(\ref{israel}), which for the metric (\ref{bhmetric}) are written as
\begin{eqnarray}\label{bhisrael}
{{[\partial_z a]} \over {a_0 ^2 c_0}} &=& -{{\kappa_{(6)} ^2} \over 4}
\rho  \, , \nonumber \\ 
{{[\partial_z n]} \over {a_0 c_0 n_0}} &=& {{\kappa_{(6)} ^2} \over 4}
(4p + 3 \rho) \, , 
\end{eqnarray}
where here 
\begin{eqnarray}\label{a-n}
[\partial_z a] &=& -{2 \over l} \, , \nonumber \\
{[\partial_z n]} &=& -{{h'(z_0)} \over {\sqrt{h(z_0)}}} \, .
\end{eqnarray}

\section{The Shortest Cut Equation}

We consider the metric (\ref{metric}) with $k=0$
\begin{equation}\label{shortmetric}
ds^2 = -n^2 (z) dt^2 + a^2(z) f^2(r) dr^2 + b^2(z) dy^2 +
d^2(z) dz^2  \, ,
\end{equation}
where the graviton path is defined equating (\ref{shortmetric}) to
zero. Therefore, 
\begin{equation}\label{gravpath}
\int_{r_0} ^r f(r') dr'= \int_{t_0} ^t {{\sqrt{n^2(z) - b^2(z)
      \dot y^2 - d^2(z) \dot z^2}}\over{a(z)}} dt \equiv \int_{t_0}
      ^t {\cal L} \left[y(t), \dot y (t), z(t), \dot z (t) ; t \right] dt
\end{equation}
which naturally defines a lagrangian density. The Euler-Lagrange
equations of ${\cal L}$ define the graviton path. We first choose to
work at a constant $y$ to check on the very possibility of
(\ref{shortmetric}) allowing shortcuts. In this case the resulting
equation is simple but far from trivial,
\begin{equation}\label{zsym}
\ddot z + \left( {{a'}\over a} -2 {{n'}\over n} + {{d'}\over d}
\right) \dot z ^2 + \left( {{n n'}\over{d^2}} - {{a'}\over a} {n^2
    \over d^2} \right) = 0 \, .
\end{equation}

For $z\leq z_0$, $a=z/l$, $n=\sqrt{h(z)}$, and $d=1/\sqrt{h(z)}$. For
$z \geq z_0$ we have to use the $Z_2$ symmetry showed up in
(\ref{z2sym}). 

Notice that this case is equivalent to consider the problem in five
dimensions with the metric shown in \cite{csaki1}.

The most general case includes a $y$ dependence on the graviton path
and the two Euler-Lagrange equations are then given by
\begin{eqnarray}\label{ynoysym}
(n^2 - d^2 \dot z ^2) \ddot y + \dot z \left\{ \left( - {{a'}\over a}
    + 2 {{b'} \over b } \right) (n^2 - d^2 \dot z ^2) -nn' + \right.&&
    \nonumber \\ 
\left.+ dd' \dot z ^2 + d^2 \ddot z \right\} \dot y + b^2 \left(
    {{a'}\over a} - {{b'} \over b } \right) \dot z \dot y ^3 = 0 &&
\end{eqnarray}
and
\begin{eqnarray}\label{znoysym}
(n^2 - b^2 \dot y ^2) \ddot z + \left\{ \left({{a'}\over a} +
    {{d'}\over d} \right) (n^2 - b^2 \dot y ^2) - 2nn' + \right. &&
    \nonumber \\  
 +2bb' \dot y ^2 \big\} \dot z ^2 +
(b^2 \dot y \ddot y) \dot z +  && \nonumber \\ +\left\{ -{{a'}\over{ad^2}} (n^2 - b^2 \dot y ^2) + {{nn' - bb' \dot y
    ^2}\over d^2} \right\} (n^2 - b^2 \dot y ^2) =0 \, . &&
\end{eqnarray}

It is clear that the case $\dot y =0$ is a solution of this set of equations
when at the same time $z$ obeys (\ref{zsym}).

This set of equations can be handled leading to 
\begin{eqnarray}\label{geosys}
{{z \dot z \dot y}\over {h(z)}} F_z + \left( h(z) - {{\dot z^2}
    \over {h(z)}} \right) F_y &=& 0 \, ,\nonumber \\
\left( 1- {{z^2 \dot y^2} \over {h(z)}}\right) F_z +{{z \dot z \dot
    y}\over {h(z)}} F_y &=& 0 \, ,
\end{eqnarray}
where
\begin{eqnarray}\label{geodesic}
F_y &=& \ddot y + \dot z \left( {2 \over z} - {{h'(z)}\over{h(z)}}
\right) \dot y \, , \nonumber \\
F_z &=& {{\ddot z} \over z} + {{\dot z^2} \over {z^2}} \left( 1- {3
    \over 2} z  {{h'(z)}\over{h(z)}}\right) + {{h(z)}\over z}
\left({{h'(z)}\over 2} - {{h(z)}\over z} \right) \, .
\end{eqnarray}
Since the determinant of the set (\ref{geosys}) is non-zero, the
solutions of (\ref{ynoysym}) and (\ref{znoysym}) must satisfy $F_y=0$
and $F_z=0$ independently. Furthermore, let us notice that $F_y=0$
and $F_z=0$ are the null geodesic equations for $y$ and $z$
respectively obtained from  
\begin{equation}\label{geoeq}
\ddot x ^\alpha + \Gamma^\alpha _{\mu\nu} \dot x^\mu \dot x^\nu =
\lambda \dot x^\alpha \, .
\end{equation}
Thus, a null curve is extreme if and only if it is a null geodesic.

Then, our problem is reduced to the previous case with constant
$y$ described by (\ref{zsym}).

For $k \not= 0$ cases we can also consider (\ref{zsym}) as the
shortcut equation if we assume the existence of a $y$-symmetry in our
problem. 

\section{AdS-Schwarzschild Bulk}

From the Israel conditions (\ref{bhisrael}) together with (\ref{a-n})
we have 
\begin{eqnarray}
{h \over {z_0 ^2}} &=& {{\kappa_{(6)} ^4 \rho ^2} \over {64}} \, ,
\label{jumpsch1} \\
{{h'}\over{2z_0}} &=& -{{\kappa_{(6)} ^4 \rho ^2} \over {64}}
(4\omega+3) \, , \label{jumpsch2}
\end{eqnarray}
and we can obtain the black hole mass $M$ as a function of the brane
energy density $\rho$, while $\rho$ is fixed by a {\it fine-tunning},
\begin{eqnarray}
{M\over{z_0 ^5}}&=& {2 \over 5} {k \over {z_0 ^2}} - (\omega +1)
{{\kappa_{(6)} ^2 \rho ^2} \over {40}} \, , \label{6DMQAdSSch1} \\
{{\kappa_{(6)} ^2 \rho ^2} \over {64}}&=& -{{3k}\over{z_0 ^2 (8\omega
    +3)}} - {5 \over {(8\omega +3) l^2}} \, , \label{6DMQAdSSch2}
\end{eqnarray}
where $\omega=p/\rho$.

\begin{figure*}[htb!]
\begin{center}
\leavevmode
\begin{eqnarray}
\epsfxsize= 7.0truecm\rotatebox{-90}
{\epsfbox{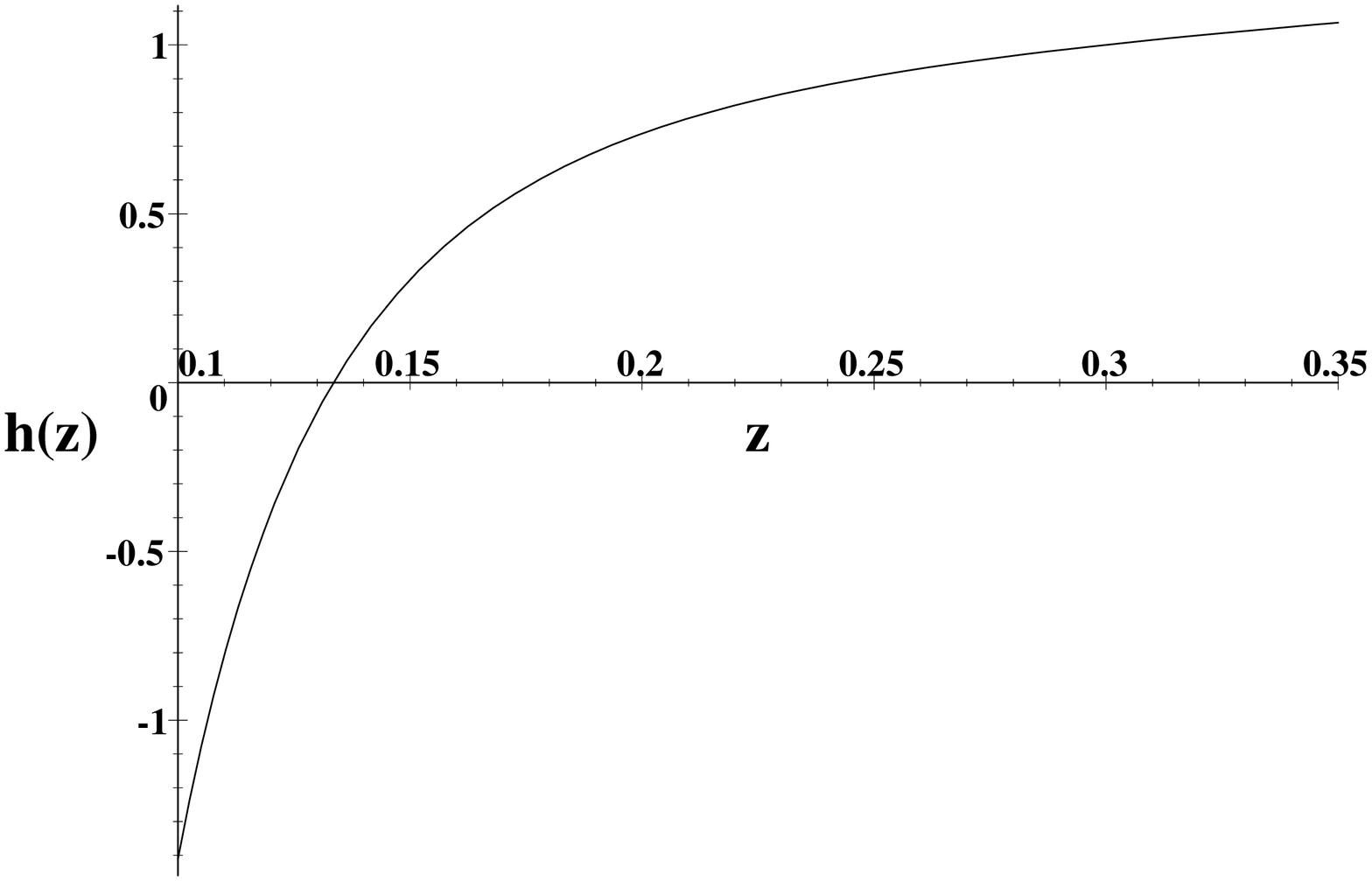}}\nonumber
\end{eqnarray}
\caption{$h(z)$ in six-dimensional AdS-Schwarzschild bulk with the
 brane located at $z=1/3$. Notice that the singularity is shielded by a
 horizon.}  
\label{6DAdSSch1}
\end{center}
\end{figure*}

\begin{figure*}[htb!]
\begin{center}
\leavevmode
\begin{eqnarray}
\epsfxsize= 7.0truecm\rotatebox{-90}
{\epsfbox{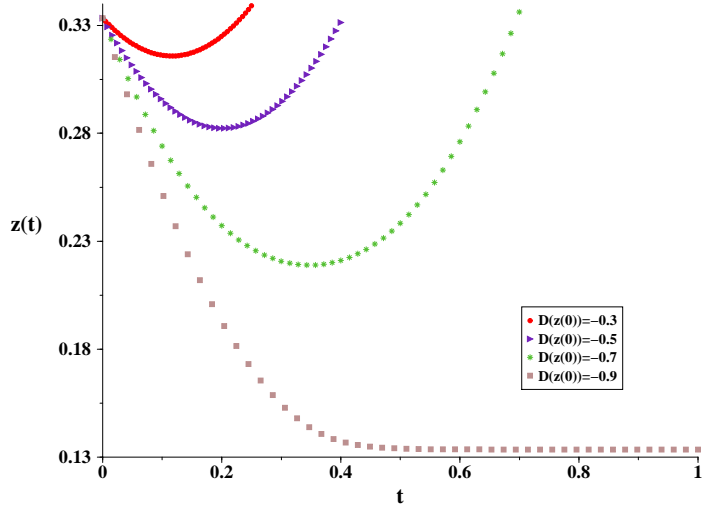}}\nonumber 
\end{eqnarray}
\caption{Shortcuts for several initial velocities in six-dimensional
 AdS-Schwarzschild bulk. Notice that there is a threshold initial velocity for
 which the graviton can not return to the brane and falls into the
 event horizon.}   
\label{6DAdSSch2}
\end{center}
\end{figure*}

\begin{figure*}[htb!]
\begin{center}
\leavevmode
\begin{eqnarray}
\epsfxsize= 7.0truecm\rotatebox{-90}
{\epsfbox{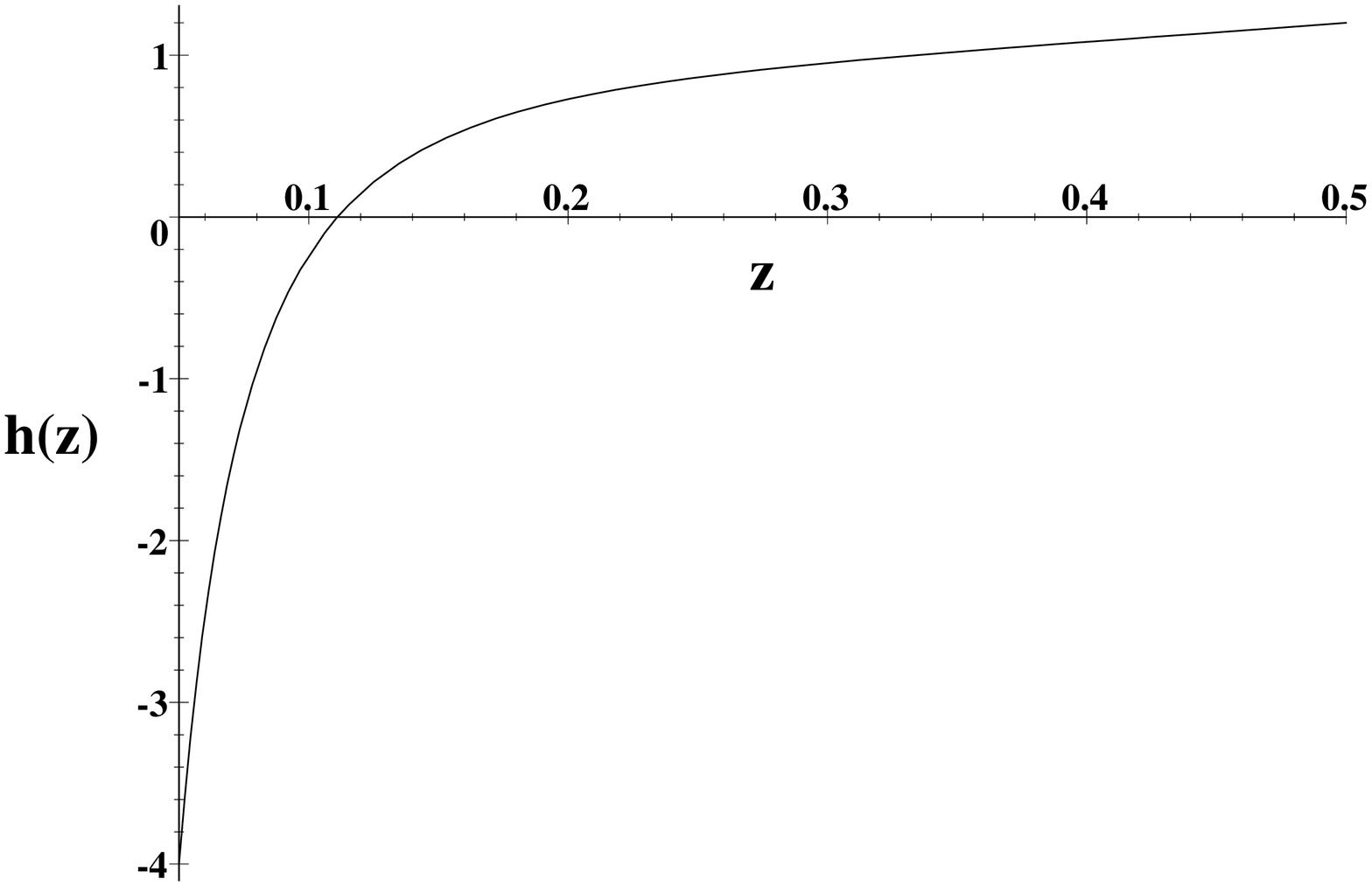}}\nonumber
\end{eqnarray}
\caption{$h(z)$ in five-dimensional AdS-Schwarzschild bulk with the
 brane located at $z=1/2$. Notice that the singularity is shielded by a
 horizon.} 
\label{5DAdSSch1}
\end{center}
\end{figure*}

\begin{figure*}[htb!]
\begin{center}
\leavevmode
\begin{eqnarray}
\epsfxsize= 7.0truecm\rotatebox{-90}
{\epsfbox{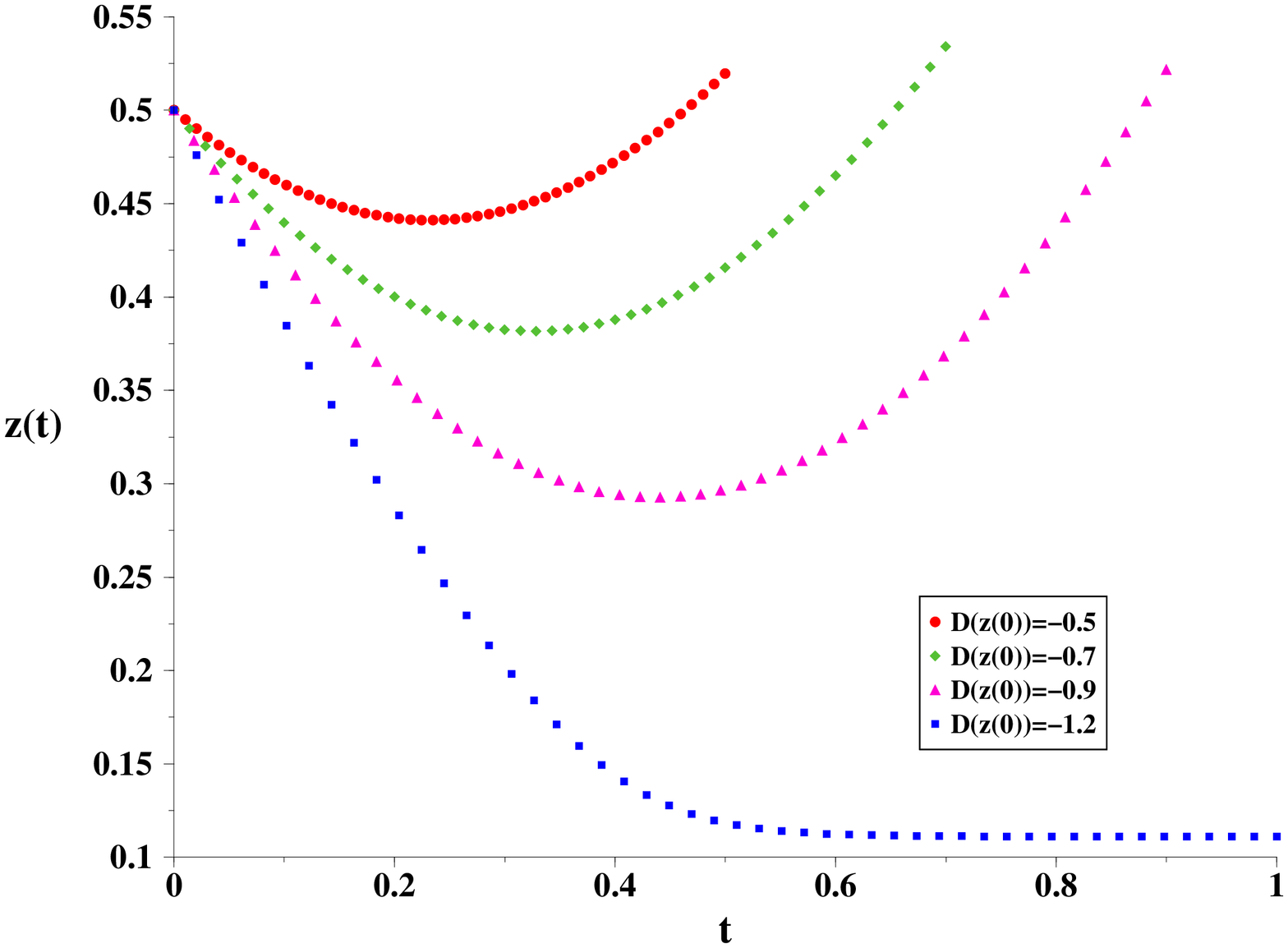}}\nonumber
\end{eqnarray}
\caption{Shortcuts for several initial velocities in five-dimensional
 AdS-Schwarzschild bulk. As in the six dimensional case, there is a
 threshold initial velocity for which the graviton can not return to the brane
 and falls into the event horizon.}  
\label{5DAdSSch2}
\end{center}
\end{figure*}


As we saw in the previous section, the shortcuts in six dimensions are
determined from (\ref{zsym}). We should also remember that the
brane is static at $z=z_0$. 

If a shortcut exists, there must be a time $t=v$ in the graviton path
when $\dot z (v)=0$ and $\ddot z(v) \geq 0$. Thus, (\ref{zsym})
evaluated at this point will give
\begin{equation}\label{zv}
\ddot z (v) + h(z_v) \left( {{h'(z_v)}\over 2} - {{h(z_v)}\over {z_v}}
\right) =0 \, .
\end{equation}

It is obvious that this minimum must be between the brane and the
event horizon $z_h$, if a horizon exists. Otherwise, there is no turning
point in the path since the graviton can not return after it goes
through the event horizon. Hence, $h(z_v) >0$. 

Thus, from (\ref{zv}) we require
\begin{equation}\label{F}
F(z_v) = {{h'(z_v)}\over 2} - {{h(z_v)}\over {z_v}} \leq 0 \quad
\hbox{for} \quad z_{h} < z_v < z_0 \, .
\end{equation}
Using (\ref{hsch}) this implies 
\begin{equation}\label{Fsch}
F(z) = {5\over 2} {M\over z^4} -{k \over z} \, .
\end{equation}
This equation has a zero in $z=z_f \not= 0$ for $k\not= 0$
$$ z_f ^3= {5 \over{2k}} M \, .$$ Thus, for the $k=0$ or $k=-1$ cases
there is no positive root. Since the mass, $M$, is positive, $F(z)>0$
everywhere preventing the coexistence of shortcuts and horizons.

On the other hand, for $k=1$ there is one real and positive root,
which must satisfy $z_f< z_0$ in order to have shortcuts. This is
$${{5M}\over {2z_0^3}} -1 <0 \, .$$
Taking into account (\ref{6DMQAdSSch1}) and the fact that
$\varepsilon^2$ must be positive in (\ref{6DMQAdSSch2}) \footnote{From
now on, we will denote $\varepsilon^2=\kappa_{(6)} ^4 \rho^2 /64$ in
six dimensions.}
\begin{equation}\label{schom1}
-4(\omega +1) \varepsilon^2 z_0 ^2 <0 \, ,
\end{equation}
then
\begin{equation}\label{schom2}
\omega +1 >0 \, .
\end{equation}

Now, let us study the conditions under which the event horizon must
appear. In general, the horizons occur at the zeros of $h(z)$, or
equivalently at the zeros of
$$ z^5+z^3 l^2 -l^2 M \, .$$
In the meantime, the non-vanishing zeros of $h'(z)$ occur when
$$ 2z^5 + 3l^2 M=0 \, .$$
Since the derivative has no positive zeros with $M>0$, there is just
one event horizon. Then as $h(z)$ goes to $-\infty$ at the origin, the
conditions
\begin{equation}
M>0 \label{schcond1}
\end{equation}
and 
\begin{equation}
h(z_0)>0 \label{schcond2}
\end{equation}
are necessary and, in fact,
enough to have a horizon and assure that the brane lies after it.

The condition (\ref{schcond2}) is automatically satisfied due to
equation (\ref{jumpsch1}).

To fulfill (\ref{schcond1}) let us substitute (\ref{6DMQAdSSch2}) into
(\ref{6DMQAdSSch1}) to have
$$ (\omega + {3\over 4}) + (\omega +1) {{z_0 ^2}\over l^2} <0 \, .$$
If $\omega +1 \leq 0$, this condition is always satisfied, but this
configuration does not produce shortcuts as we would like. However, the
condition is also satisfied with $\omega +1 >0$ if we require
\begin{equation}\label{schomega}
-1<\omega < -{3\over 4} \, ,
\end{equation}
and
\begin{equation}\label{schbpos}
{{z_0 ^2}\over l^2} < - {{\omega + 3/4}\over {\omega +1 }} \, .
\end{equation}

If we follow both (\ref{schomega}) and (\ref{schbpos}) together with
the fine-tunning for the energy (\ref{6DMQAdSSch2}), we will have several
shortcuts in AdS-Schwarzschild bulks with shielded singularity.
In figures \ref{6DAdSSch1} and \ref{6DAdSSch2} we illustrate an example
with $\omega=-4/5$, $z_0=1/3$, and $l=1$. Notice in figure
\ref{6DAdSSch1} that the horizon appears before the brane. 

Since this case is equivalent to consider the problem in five
dimensions with $h(z)$, $M$ and $\rho$ given in \cite{csaki1},
analogous results are obtained. In this case, the fine-tunning in the
energy is given by \footnote{In this case
$\varepsilon_{(5)} ^2=\kappa_{(5)} ^4 \rho^2 /36$.} 
\begin{equation}\label{5Dfine}
\varepsilon_{(5)} ^2=-{1\over{3\omega+1}} \left({{1}\over {z_0 ^2}} +
{2\over l^2} \right) \, ,
\end{equation}
and $\omega$ is confined to
\begin{equation}\label{5Dschomega}
-1<\omega < - {2\over 3} \, ,
\end{equation}
while the brane position is given by
\begin{equation}\label{5Dbpos}
{{z_0 ^2}\over l^2} < - {{\omega + 2/3}\over{\omega +1}} \, .
\end{equation}

An example is shown in figures \ref{5DAdSSch1} and \ref{5DAdSSch2} for
$\omega=-3/4$, $z_0=1/2$, and $l=1$.

\section{AdS-Reissner-Nordstr\"om Bulk}

From the Israel conditions (\ref{bhisrael}) we will have for the black
hole mass and charge,
\begin{eqnarray}\label{6DMQAdSRN}
{M\over {z_0 ^5}} &=& {{2k}\over{z_0 ^2}} + {8\over{3l^2}} +
{{\kappa_{(6)} ^4} \over {24}} \rho^2 \omega \, , \nonumber \\
{Q^2 \over {z_0 ^8}} &=& {k\over{z_0 ^2}} + {5 \over{3l^2}} +
{{8\omega +3}\over 3} \, {{\kappa_{(6)} ^4 \rho^2} \over {64}} \, . 
\end{eqnarray}

At this stage it is convenient to carefully study the possibility
of existence of shortcuts for every value of $k$. 

\subsection{$k=0$ and $k=-1$ Cases}

As it was found in the AdS-Schwarzschild case, (\ref{F}) determines
the existence of shortcuts.
Using (\ref{hrn}) we see that (\ref{F}) has a zero in $z=z_f\not=0$
when
\begin{equation}\label{zerosF}
{5\over 2} M z_f ^3 -4Q^2 -kz_f ^6=0 \, .
\end{equation}
If $k=0$, we have a real root in
\begin{equation}\label{k0root}
z_f ^3 = {{8Q}\over{5M}} \, .
\end{equation}
If $k=1$, we have two roots in
\begin{equation}\label{k1root}
z_f ^3 = {5\over4}M \pm {1\over 4} \sqrt{25M^2-64Q^2} \, .
\end{equation}
Finally, if $k=-1$, we have
\begin{equation}\label{k-1root}
z_f ^3 = -{5\over4}M \pm {1\over 4} \sqrt{25M^2+64Q^2} \, .
\end{equation}

Notice that $F(z)$ has at most one real and positive zero if $k=0, -1$
and at most two positive zeros if $k=1$.

Analyzing $h(z)$ and its derivative we see that $h(z)$ tends to
$+\infty$ both at the singularity and at infinity, while $h'(z)$ tends to
$-\infty$ at the singularity and to $+\infty$ at infinity.

The horizons occur at the zeros of $h(z)$, or equivalently, at the
zeros of
\begin{equation}\label{polz}
z^8+l^2 k z^6 - l^2 M z^3 + l^2Q^2 =0\, .
\end{equation}

On the other hand, the non-vanishing zeros of $h'(z)$ occur when
\begin{equation}\label{dpolz}
2 z^8 +3l^2M z^3 -6l^2 Q^2 =0 \, .
\end{equation}
This polynom grows at infinity being negative at the
origin. Its derivative has non-vanishing roots when
\begin{equation}\label{ddpolz}
16z^5 +9l^2 M =0 \, .
\end{equation}
For $M>0$ this equation is never satisfied. Thus, as the derivative of
(\ref{dpolz}) does not vanish and is positive outside the origin,
the polynom (\ref{dpolz}) grows monotonically and has just one
root. The zeros of this polynom are all non-vanishing zeros of
$h'(z)$. Therefore, we conclude that for positive mass there is just
one zero for $h'(z)$, and hence, at most two horizons for $h(z)$.

When there is one horizon, $h'(z)$ is negative before it and positive
after, crossing $h(z)$ at the very horizon. If there are two horizons,
$h'(z)$ vanishes at a point between Cauchy and event horizons,
being negative before this point and positive after, while $h(z)$ is
positive at all points except between both horizons. Taking into
account both the sign and zeros of these functions, $h'(z)$ crosses
$h(z)$ between the Cauchy horizon and the point at which $h'(z)$ vanishes.

Since $h'(z)/2$ has the same sign as $h'(z)$ and vanishes at the same
point, and in the same way $h(z)/z$ has the same sign of $h(z)$ and
vanishes at the same points, we conclude that, existing horizons, $F(z)$
necessarily vanishes at some point $z=z_c$ such that $0<z_c<z_h$. 
However, as we pointed out before, for $k=0$ or $k=-1$ there is only
one positive root of $F(z)$. As $F(z)<0$ for $z<z_c$, then $F(z)>0$
for $z>z_c$. Thus, because $z_c \leq z_h$, $F(z)>0$ for $z>z_h$
contrary to what was required in (\ref{F}). This implies that
there are no shortcuts with $k=0$ or $k=-1$ when horizons exist. 

In five dimensions the proof is very similar and we arrive to the same
conclusion.

\subsection{$k=1$ Case}

As we saw in the previous section, $F(z)$ has two real, positive and
distinct roots for $k=1$,
\begin{eqnarray}
r_1 ^3 = {5\over4}M - {1\over 4} \sqrt{25M^2-64Q^2} \, , \label{root1}\\
r_2 ^3 = {5\over4}M + {1\over 4} \sqrt{25M^2-64Q^2} \, . \label{root2}
\end{eqnarray}
This is the only situation where the shortcuts can coexist with a
shielded singularity. In fact, this situation necessarily requires
the second root of $F(z)$ being at some point before the brane
position $z_0$. This also implies $F(z_0)<0$.

In addition, we must have both $Q^2$ and $M$ positive.

Given the fact that we have horizons, if the brane is not between them
or at a horizon 
position, then $h(z_0)>0$. Furthermore, in order to guarantee that the brane
is located after the event horizon, we also need $h'(z_0)>0$.

From the discussion in the previous section we will have one or two
horizons if and only if $h(r_1)\leq 0$.

In summary, shortcuts in bulks with shielded singularities can occur
only if $k=1$ and also if the following conditions are supplied,
\begin{enumerate}
\item $h(z_0)>0$ and $h'(z_0)>0$ to have both horizons before the
brane.

\item $F(z_0)<0$ and $r_2<z_0$ to have shortcuts with shielded
singularity.

\item $Q^2>0$ and $M>0$, which assures the positivity of the black
hole mass and charge.

\item $h(r_1)\leq 0$ in order to have horizons.
\end{enumerate}

We will analyze each condition and impose certain restrictions on
$\omega$, $\rho^2$, and $z_0$.

\begin{subsubsection}{Existence of Both Horizons Before the Brane}
These conditions are the simplest to analyze since they restrict
$\omega$ directly from the Israel conditions (\ref{bhisrael}) together
with (\ref{a-n})
\begin{eqnarray}
{{h(z_0)}\over{a_0 ^2}} &=& \varepsilon^2 \, ,\label{jumpb} \\
{{h'(z_0)}\over{2z_0}} &=& -(4\omega +3) \varepsilon^2 \, . \label{jump}
\end{eqnarray}
The condition (\ref{jumpb}) is automatically satisfied since
$\varepsilon^2>0$. 

From the condition (\ref{jump})
\begin{equation}\label{jump1}
-(4\omega +3) \varepsilon^2 >0 \, ,
\end{equation}
we have our first restriction
\begin{equation}\label{ome1}
\omega < -3/4 \, .
\end{equation}
\end{subsubsection}

\begin{subsubsection}{Existence of Shortcuts with Shielded Singularity}
From the definition of $F(z)$, (\ref{F}), and using (\ref{jumpb}) and
(\ref{jump}) we see that 
\begin{equation}\label{Fz0}
0> F(z_0) = -4(\omega+1) \varepsilon^2 z_0 \, .
\end{equation}
Thus we find another condition on $\omega$
\begin{equation}\label{ome2}
\omega +1 >0 \, .
\end{equation}

Besides, from $r_2<z_0$
\begin{equation}\label{r2<z0}
{{5M}\over 4} - z_0 ^3 < -{1\over 4} \sqrt{25M^2 -64 Q^2} \, .
\end{equation}
This equation will be satisfied if \footnote{We assume that $25M^2 -64
Q^2 >0$. We will return to this condition when we discuss the
existence of horizons, where we will impose a stronger restriction, $M^2
-4Q^2 >0$.}
\begin{equation}\label{r2<z0.2}
{{5M}\over 4} -z_0 ^3 < 0 \, ,
\end{equation}
or using (\ref{6DMQAdSRN})
\begin{equation}\label{r2<z0.3}
{3 \over 2}z_0 ^3 +{{10}\over 3} z_0 ^5 + {{10}\over 3} z_0 ^5 \omega
\varepsilon^2 <0 \, ,
\end{equation}
and as $\omega <-3/4$
\begin{equation}\label{ep1}
z_0 ^2 \varepsilon^2 > {1 \over \omega} \left( -{9\over {20}} - {{z_0
^2}\over l^2} \right) \, .
\end{equation}
\end{subsubsection}

\begin{subsubsection}{Positivity of the Black Hole Mass and Charge}
Because we require the positivity of the black hole mass, from
(\ref{6DMQAdSRN}) we have
\begin{equation}\label{M>0}
{M \over {z_0 ^3}} > 0 \quad \Rightarrow \quad {{z_0 ^2}\over l^2} + \omega
\varepsilon^2 z_0 ^2 > -{3\over 4} \, ,
\end{equation}
thus,
\begin{equation}\label{M>0.2}
z_0 ^2 \varepsilon^2  < {1\over \omega} \left(-{3\over 4} - {{z_0
^2}\over l^2} \right) \, .
\end{equation}
Since $3/4>9/20$ this condition is certainly compatible with
(\ref{ep1}).

On the other hand, the positivity of the squared black hole charge
requires
\begin{equation}\label{Q>0}
{Q^2 \over {z_0 ^6}} >0 \quad \Rightarrow \quad 1+{{5z_0 ^2} \over {3l^2}} +
\left( {8\over 3}\omega +1 \right) z_0 ^2 \varepsilon^2 >0 \, ,
\end{equation} 
so that
\begin{equation}\label{Q>0.2}
z_0 ^2 \varepsilon^2 < {1 \over {8\omega +3}} \left(-3 -{{5z_0
^2}\over l^2} \right) \, .
\end{equation}
In spite of not being trivial, this equation is also compatible with
(\ref{ep1}). This requires
\begin{equation}\label{comp1}
{1\over \omega} \left( -{9\over {20}} - {{z_0 ^2}\over l^2} \right) <
{1 \over {8\omega +3}} \left(-3 -{{5z_0 ^2}\over l^2} \right) \, ,
\end{equation}
or 
\begin{equation}\label{comp2}
-{1\over 5} \left( \omega + {9\over 4} \right) - {{z_0 ^2}\over l^2}
\left(\omega +1 \right) <0 \, ,
\end{equation}
what is always true for $-1<\omega <-3/4$.
\end{subsubsection}

\begin{subsubsection}{Existence of Horizons}
This is the last and the more complicated of our conditions. We must
have $h(r_1)\leq 0$. Let $x$ be $r_1 ^3$,
\begin{equation}\label{x}
{{x^{8/3}}\over l^2} + x^2 -Mx + Q^2 \leq 0 \, .
\end{equation}
We do not need to do a complete study of this equation. For our
purposes it will be enough to require
\begin{equation}\label{x2}
x^2-Mx +Q^2 <0 \, .
\end{equation} 
Using (\ref{root1}) this implies
\begin{equation}\label{MQ1}
M^2-4Q^2 >0 \, .
\end{equation}
This condition is necessary but not enough to have horizons. However,
this restriction added to the others developed in this section will be
enough to construct shortcuts with horizons as we will see. Notice
that this condition is stronger than that one assumed before, $M>8/5
\, Q$. 

Using (\ref{6DMQAdSRN}) in (\ref{MQ1}),
\begin{equation}\label{MQ2}
(1-\varepsilon^2 l^2) + {{16}\over 9} {{z_0 ^2} \over l^2}
\left(1+\omega \varepsilon^2 l^2 \right)^2 >0 \, .
\end{equation}

We know from (\ref{ep1}) that $1-\varepsilon^2 l^2$ must be
negative. Therefore, we must very carefully analyze (\ref{MQ2}). 
We can interpret (\ref{MQ2}) as a quadratic equation in the energy
\begin{equation}\label{epsqr}
\left(1+{{16}\over 9} {{z_0 ^2}\over l^2} \right) + \left( {{32}\over
9} \omega z_0 ^2 -l^2 \right) \varepsilon^2 + \left( {{16}\over 9} l^2
z_0 ^2 \omega ^2 \right) \varepsilon^4 >0 \, ,
\end{equation}
what implies
\begin{equation}\label{ep2}
z_0 ^2 \varepsilon^2 > {1 \over {32}} \left( -32 \omega z_0 ^2 + 9l^2
+ 3 \sqrt{-64\omega z_0 ^2 l^2 + 9l^4 - 64 l^2 z_0 ^2 \omega ^2}
\right) / (\omega^2 l^2) \, ,
\end{equation}
or 
\begin{equation}\label{ep3}
z_0 ^2 \varepsilon^2 < {1 \over {32}} \left( -32 \omega z_0 ^2 + 9l^2
- 3 \sqrt{-64\omega z_0 ^2 l^2 + 9l^4 - 64 l^2 z_0 ^2 \omega ^2}
\right) / (\omega^2 l^2) \, .
\end{equation}
Notice that because $-1<\omega <-3/4$, $$-64\omega z_0 ^2 l^2 -64 l^2
z_0 ^2 \omega ^2 = -64 \omega (\omega +1) z_0 ^2 l^2 >0 $$ and all the
previous roots are real and positive.
\end{subsubsection}

\bigskip

Summarizing, from the considerations in the previous sections, we must
have for $\omega$
\begin{equation}\label{om}
-1 <\omega < -3/4 \, .
\end{equation}
For the energy,
\begin{eqnarray}
&{1 \over \omega} \left( -{9\over {20}} - {{z_0 ^2}\over l^2} \right) <
z_0 ^2 \varepsilon^2 < {1\over \omega} \left(-{3\over 4} - {{z_0
^2}\over l^2} \right) \, , & \quad \hbox{or} \label{eps1} \\
&{1 \over \omega} \left( -{9\over {20}} - {{z_0 ^2}\over l^2} \right) <
z_0 ^2 \varepsilon^2 < {1 \over {8\omega +3}} \left(-3 -{{5z_0
^2}\over l^2} \right) \, , & \label{eps2}
\end{eqnarray}
depending on which condition is more restrictive.

In addition,
\begin{eqnarray}
&z_0 ^2 \varepsilon^2 > {1 \over {32}} \left( -32 \omega z_0 ^2 + 9l^2
+ 3 \sqrt{-64\omega z_0 ^2 l^2 + 9l^4 - 64 l^2 z_0 ^2 \omega ^2}
\right) / (\omega^2 l^2) \, , & \label{eps3} \\ \hbox{or} \nonumber \\
&z_0 ^2 \varepsilon^2 < {1 \over {32}} \left( -32 \omega z_0 ^2 + 9l^2
- 3 \sqrt{-64\omega z_0 ^2 l^2 + 9l^4 - 64 l^2 z_0 ^2 \omega ^2}
\right) / (\omega^2 l^2) \, . & \label{eps4}
\end{eqnarray}

Now we are going to analyze the situations in which all these
conditions are compatible.

Let us begin our analysis with equation (\ref{eps4}). To be compatible
with (\ref{eps1}) and (\ref{eps2}), we just need
\begin{equation}\label{compat4-12}
{1 \over \omega} \left( -{9\over {20}} - {{z_0 ^2}\over l^2} \right) <
{1 \over {32}} \left( -32 \omega z_0 ^2 + 9l^2
- 3 \sqrt{-64\omega z_0 ^2 l^2 + 9l^4 - 64 l^2 z_0 ^2 \omega ^2}
\right) / (\omega^2 l^2) \, ,
\end{equation}
that is,
\begin{equation}\label{compat4-12.2}
- {9 \over {20}} \omega - {9\over {32}} + {3\over {32}} \sqrt{-64
\omega {{z_0 ^2}\over l^2} +9 -64{{z_0 ^2}\over l^2} \omega ^2} <0 \, .
\end{equation}
Since $3/4 <\mid \omega \mid <1$, $$-{9\over {20}} \omega -{9\over
{32}}$$ will always be positive and thus, (\ref{compat4-12.2}) will never be
satisfied. Then, we conclude that (\ref{eps4}) is not compatible
either with (\ref{eps1}) or (\ref{eps2}). This implies that $z_0 ^2
\varepsilon^2$ must satisfy (\ref{eps3}) together with (\ref{eps1}) or
(\ref{eps2}).

Let us initially compare (\ref{eps1}) with (\ref{eps3}). We must have
\begin{equation}\label{compat3-12}
{1\over \omega} \left(-{3\over 4} - {{z_0 ^2}\over l^2} \right) > {1
\over {32}} \left( -32 \omega z_0 ^2 + 9l^2 + 3 \sqrt{-64\omega z_0 ^2
l^2 + 9l^4 - 64 l^2 z_0 ^2 \omega ^2} \right) / (\omega^2 l^2) \, ,
\end{equation}
that is,
\begin{equation}\label{compat3-12.2}
-{3\over 4}\omega -{9 \over {32}} - {3\over {32}} \sqrt{-64\omega
{{z_0 ^2}\over l^2} + 9 -64{{z_0 ^2}\over l^2} \omega^2 } > 0 \, .
\end{equation}
In this case, since $\omega$ is negative and $3/4 <\mid \omega \mid
<1$, $$-{3\over 4} \omega -{9 \over {32}} >0$$ and
(\ref{compat3-12.2}) can be satisfied if
\begin{equation}\label{compat3-12.3}
\left(-{3\over 4} \omega -{9 \over {32}}\right)^2 > {9\over{1024}}
\left(-64\omega {{z_0 ^2}\over l^2} + 9 -64{{z_0 ^2}\over l^2}
\omega^2 \right) \, , 
\end{equation}
or
\begin{equation}\label{compat3-12.4}
{9\over{16}} \omega (\omega +1) {{z_0 ^2}\over l^2} + {9 \over {64}}
(3+4\omega) \omega >0 \, .
\end{equation}
Since $\omega +1 >0$ and $3+4\omega <0$, for positive $z_0$ the
inequality will be only fulfilled if 
\begin{equation}\label{z-l}
{z_0 \over l} < {1\over 2} \sqrt{- {{3+4\omega}\over{1+\omega}}} \, .
\end{equation}
So that (\ref{eps3}) and (\ref{eps1}) can be compatible.

Now we are going to analyze the compatibility between (\ref{eps3}) and
(\ref{eps2}). We must have
\begin{equation}\label{compat3-2}
{1 \over {8\omega +3}} \left(-3 -{{5z_0 ^2}\over l^2} \right) + {{z_0
^2}\over {\omega l^2}} - {9 \over {32\omega^2}} > {3 \over
{32\omega^2}} \sqrt{-64\omega z_0 ^2 l^2 + 9l^4 - 64 l^2 z_0 ^2 \omega
^2} \, .
\end{equation}
or using that $8\omega +3<0$, simplifying, and squaring both sides we
can write (\ref{compat3-2}) as
\begin{equation}\label{compat3-2.2}
{1\over {16}} \omega^2 (3+4\omega)^2 + \omega^2 (\omega+1)^2
\left({{z_0 ^2}\over l^2} \right)^2 + {\omega^2 \over 2} (3+4\omega)
(\omega+1) {{z_0 ^2}\over l^2} >0 \, .
\end{equation}
This polynom has just one root for ${{z_0 ^2}/ l^2}$ $${{z_0
^2}\over l^2} = -{1\over 4} \left({{3+4\omega}\over {1+\omega}}\right)
\, .$$ 

Since the coeficient of $z_0 ^4/l^4$ is positive, the inequality is
satisfied with the same condition (\ref{z-l}), so we verify that
both (\ref{eps2}) and (\ref{eps1}) are compatible with (\ref{eps3})
under the same restrictions.

Furthermore, let us compare the upper limits of (\ref{eps1}) and
(\ref{eps2}). Suppose
\begin{equation}\label{supo}
{1\over \omega} \left(-{3\over 4} - {{z_0 ^2}\over l^2} \right) > {1
\over {8\omega +3}} \left(-3 -{{5z_0 ^2}\over l^2} \right) \, ,
\end{equation}
what can also be written as
\begin{equation}\label{supo.2}
-{3 \over 4} (4\omega +3) - {{3z_0 ^2}\over l^2} (\omega +1) >0 \, .
\end{equation}
Thus, (\ref{supo}) is satisfied if and only if
\begin{equation}\label{z-l.2} 
{{z_0 ^2}\over l^2} < -{1\over 4} \left({{3+4\omega}\over {1+\omega}}\right) \, ,
\end{equation} 
which is just the same inequality (\ref{z-l}), that $z_0$ must
satisfy. Hence, between (\ref{eps1}) and (\ref{eps2}), it is 
enough to take into account the latter. Nevertheless, from
(\ref{compat3-2.2}) notice that (\ref{eps2}) would be also compatible with
(\ref{eps3}) if 
$${{z_0 ^2}\over l^2} > -{1\over 4} \left({{3+4\omega}\over
{1+\omega}}\right) \, ,$$ 
and (\ref{supo}) would be satisfied with a
change of sign implying that we should consider (\ref{eps1}) instead of
(\ref{eps2}); however, as stated before, the compatibility of
(\ref{eps1}) and (\ref{eps3}) requires 
$$ {{z_0 ^2}\over l^2} < -{1\over 4} \left({{3+4\omega}\over
    {1+\omega}}\right) \, ,$$  
which contradicts our hypothesis. Therefore, the only possible configuration
is (\ref{z-l.2}).

At last, we compare the lower limits of (\ref{eps2}) and
(\ref{eps3}). Suppose
\begin{equation}\label{supoen}
{1 \over \omega} \left( -{9\over {20}} - {{z_0 ^2}\over l^2} \right)
< {1 \over {32 \omega^2 l^2}} \left( -32 \omega z_0 ^2 + 9l^2
+ 3 \sqrt{-64\omega z_0 ^2 l^2 + 9l^4 - 64 l^2 z_0 ^2 \omega ^2}
\right) \, ,
\end{equation}
or $$-{9\over{20}} \omega -{9\over {32}} < {3\over {32}}
\sqrt{-64\omega {{z_0 ^2}\over l^2} + 9 -64 {{z_0 ^2}\over
l^2}\omega^2} \, .$$ 
Squaring and simplifying we obtain
$$ {{z_0 ^2}\over l^2} > - {9 \over{20 (\omega+1)}} \left({4\over 5}
\omega +1 \right) \, .$$
For $-1<\omega<-3/4$ this inequality is always satisfied since the
right hand side is negative. Hence, we conclude that (\ref{supoen}) is
valid and between the lower limits for the energy in (\ref{eps2}) and
(\ref{eps3}), we just need to choose the latter.

\bigskip

\begin{figure*}[htb!]
\begin{center}
\leavevmode
\begin{eqnarray}
\epsfxsize= 7.0truecm\rotatebox{-90}
{\epsfbox{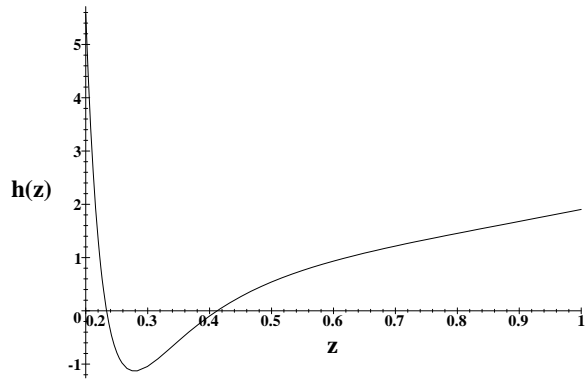}}\nonumber
\end{eqnarray}
\caption{$h(z)$ in six-dimensional AdS-Reissner-Nordstr\"om bulk with
the brane located at 
$z=1$. Notice that the singularity is shielded by two horizons.} 
\label{h(z)}
\end{center}
\end{figure*}
\begin{figure*}[htb!]
\begin{center}
\leavevmode
\begin{eqnarray}
\epsfxsize= 7.0truecm\rotatebox{-90}
{\epsfbox{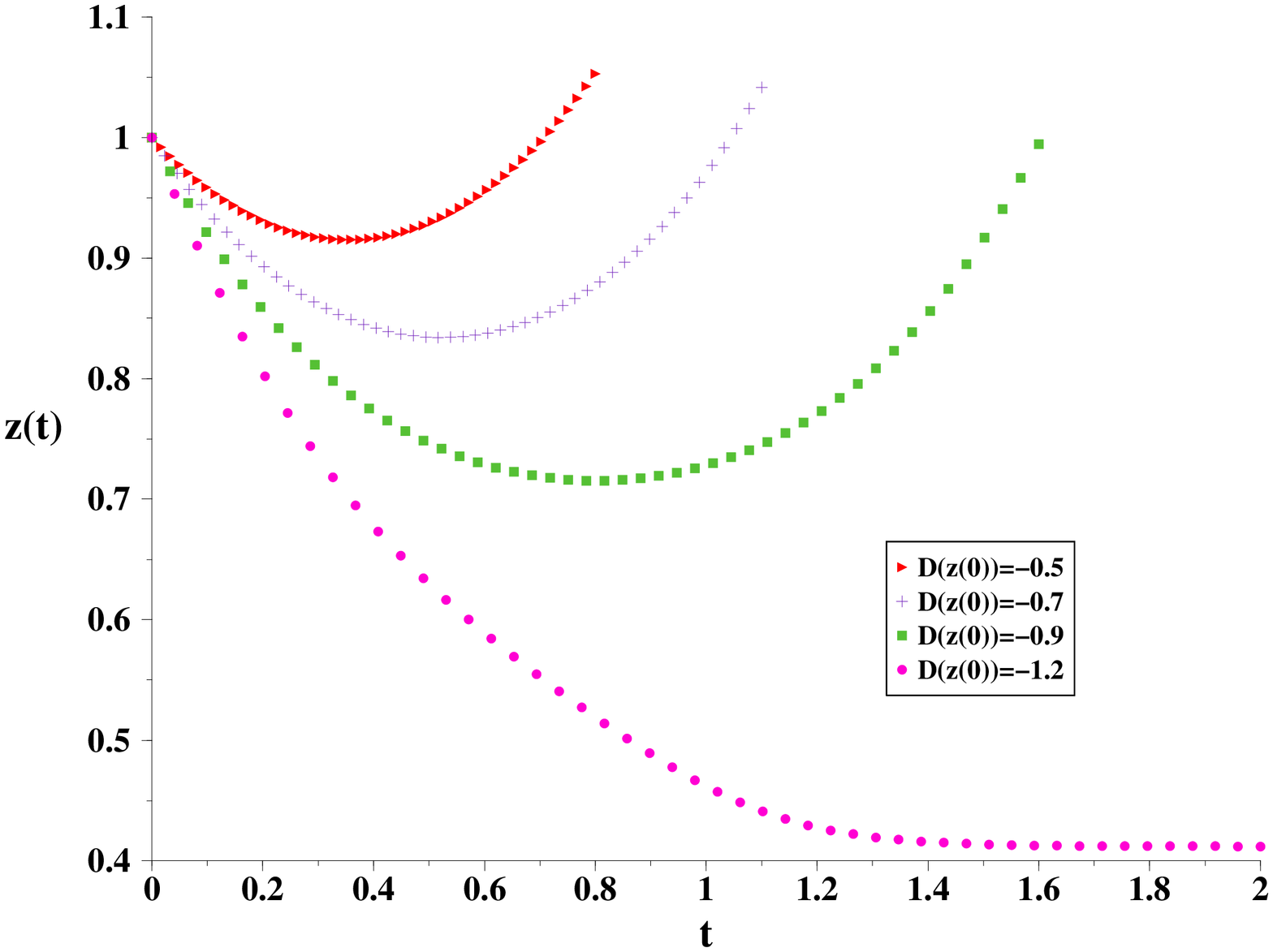}}\nonumber
\end{eqnarray}
\caption{Shortcuts for several initial velocities in six-dimensional
AdS-Reissner-Nordstr\"om bulk. Notice that there is threshold initial
velocity for which the graviton
can not return to the brane and falls into the event horizon.}  
\label{6Dsc}
\end{center}
\end{figure*}
\begin{figure*}[htb!]
\begin{center}
\leavevmode
\begin{eqnarray}
\epsfxsize= 7.0truecm\rotatebox{-90}
{\epsfbox{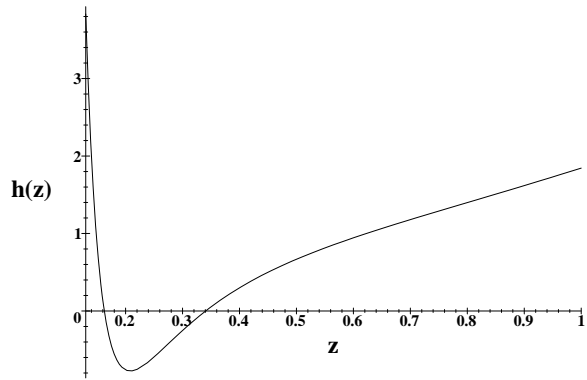}}\nonumber
\end{eqnarray}
\caption{$h(z)$ in five-dimensional AdS-Reissner-Nordstr\"om bulk with
the brane located at $z=1$. We 
see that the singularity is protected by two horizons.} 
\label{5h(z)}
\end{center}
\end{figure*}
\begin{figure*}[htb!]
\begin{center}
\leavevmode
\begin{eqnarray}
\epsfxsize= 7.0truecm\rotatebox{-90}
{\epsfbox{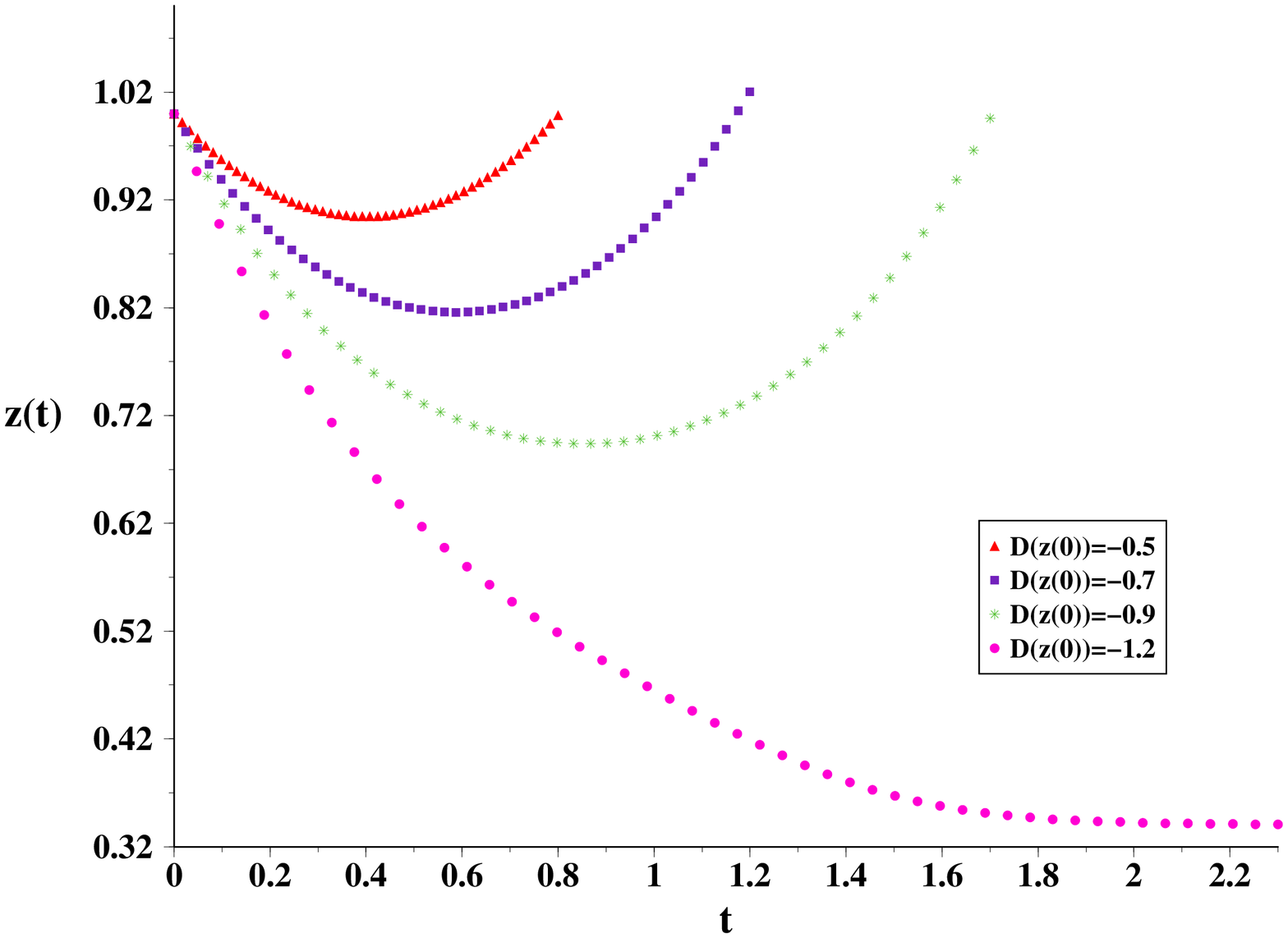}}\nonumber
\end{eqnarray}
\caption{Shortcuts for several initial velocities in five-dimensional
AdS-Reissner-Nordstr\"om bulk. We see that there is threshold initial
velocity for which the graviton
can not return to the brane and falls into the event horizon.}
\label{5Dsc}
\end{center}
\end{figure*}

In short, by purely analytic considerations we conclude that shortcuts
in bulks having no naked singularities and a static brane embedded in
can only appear if $k=1$ and if the following conditions are
satisfied,
\begin{enumerate}
\item We must choose $\omega$ such that $-1<\omega<-3/4$;

\item Given $\omega$, the brane must be located at a position such that
\begin{equation}\label{branepos}
{z_0 \over l} < {1 \over 2} \sqrt{- {{3+4\omega}\over {1+\omega}}} \, ,
\end{equation}
which is the same condition as AdS-Schwarzschild case (\ref{schbpos});

\item Given (\ref{branepos}), the energy $\varepsilon$ must satisfy
\end{enumerate}
\begin{equation}\label{energycond}
{1\over{32\omega^2}} \left( - {{32\omega z_0 ^2}\over l^2} + 9 +3
\sqrt{-64\omega {{z_0 ^2}\over l^2} +9 -64{{z_0 ^2}\over l^2}\omega^2}
\right) < z_0 ^2 \varepsilon^2 < {1\over {8\omega +3}} \left(-3-{{5z_0
^2}\over l^2} \right) \, .
\end{equation}

In this way, it turns out to be simple to find shortcuts in bulks with
shielded singularities.

As an example, let us choose $\omega=-9/10$. From (\ref{branepos}) we
must have $${z_0 \over l} < {\sqrt{6}\over 2} \, ,$$ then we choose $l=1$ and
$z_0=1$. 

From (\ref{energycond}) we have
$$ {{35}\over{24}} + {5\over{72}} \sqrt{41}< \varepsilon^2 < {{40}
\over{21}} \, ,$$ so we choose $\varepsilon = \sqrt{238/125}$. 

In figure (\ref{h(z)}) we plot $h(z)$ with these
conditions. Notice that the singularity is protected by an event horizon
and the brane is at $z=z_0=1$. 

In figure (\ref{6Dsc}) we plot the graviton paths obtained from
(\ref{zsym}) under the previous conditions for a variety of initial
velocities showing that, in fact, shortcuts appear when we choose the
parameters following the complete analysis shown in this section.

\bigskip

The analysis in five dimensions can be performed analogously to
six-dimensional one with 
\begin{equation}\label{5Dh}
h(z) = 1 + {z^2 \over l^2}- {M\over z^2}+ {Q^2\over z^4} \, ,
\end{equation}
and
\begin{eqnarray}\label{5DMQAdSRN}
{M\over {z_0 ^4}} &=& {2 \over {z_0 ^2}} + {3\over l^2} + 3\omega
\varepsilon_{(5)} ^2 \, , \\
{Q^2 \over {z_0 ^6}} &=& {1 \over {z_0 ^2}} + {2\over l^2} +3\omega
\varepsilon_{(5)} ^2 + \varepsilon_{(5)} ^2 \, ,
\end{eqnarray}
arriving to the following restrictions:

\begin{enumerate}
\item We must choose $\omega$ such that $-1<\omega<-2/3$ ;

\item  Given $\omega$, the brane must be located at a position such that
\begin{equation}\label{5Dbranepos}
{z_0 \over l} < \sqrt{- {{\omega+2/3}\over {1+\omega}}} \, ;
\end{equation}

\item Given (\ref{5Dbranepos}), the energy
$\varepsilon_{(5)} ^2=\kappa_{(5)} ^4 \rho^2 /36$ must satisfy
\end{enumerate}
\begin{equation}\label{5Denergycond}
{1\over {9 \omega^2}} \left( 2-9{{z_0 ^2}\over l^2} \omega +2
\sqrt{1-9 {{z_0 ^2}\over l^2} \omega (\omega+1)} \right) < z_0 ^2
\varepsilon_{(5)} ^2 < {1\over{3\omega +1}} \left(-1-{{2z_0 ^2}\over
l^2} \right) \, .
\end{equation}

As an example, we choose $\omega=-7/8$. From (\ref{5Dbranepos}) we
must have $${z_0 \over l} <{\sqrt{15}\over 3}\, ,$$ so we choose $l=1$ and
$z_0=1$.

From (\ref{5Denergycond}) the energy must fulfill
$$ {{632}\over{441}} + {{16}\over{441}} \sqrt{127} < \varepsilon_{(5)} ^2 <
{{24}\over {13}} \, ,$$ then we choose $\varepsilon_{(5)} =\sqrt{461/250}$.

In figure (\ref{5h(z)}) we can see $h(z)$ according to the previous
conditions. As in the six dimensional case, the singularity is
protected by an event horizon, and the brane is at $z=z_0=1$.  

In figure (\ref{5Dsc}) we show several graviton paths obtained under
the previous conditions for several initial velocities showing
that shortcuts appear when we choose the parameters
following the analysis shown in this section analogously to
the six dimensional case. 

\section{Conclusions}

In this paper we have shown that the shortest graviton path is
governed by just one equation involving the ``radial'' extra
coordinate. We have also seen that a symmetry in the ``angular'' extra
coordinate has permitted us to consider curved spatial sections. 

The AdS-Schwarzschild and AdS-Reissner-Nordstr\"om bulks open up the
possibility of having 
shortcuts provided both the spatial section has positive curvature and a
set of strong restrictions on the brane intrinsic tension must be satisfied.
Moreover, its location in the bulk has to be respected. We should also
notice that the energy is already fine-tunned from the
Israel conditions in the case of the AdS-Schwarzschild bulk .

Hence, it is interesting to notice that despite the fact that the charge
contributes 
to have a negative $F(z_0)$ and thus facilitates the existence of
shortcuts, there are more restrictive conditions for the energy coming
from $Q^2>0$ and from the horizons equation (\ref{x}) which do not
appear in the uncharged case. In this way, the results favor the existence of
shortcuts in bulks with shielded singularities with the same conditions for
$\omega$ and $z_0$ as the AdS-Schwarzschild case and also impose what
is basically a fine-tunning in the energy. Thus, both cases seem to
be equivalent for the study of shortcuts in static universes with
protected singularities. We should realize that the restrictions to
obtain these shortcuts make lose the main advantage we should have
when we study the AdS-Reissner-Nordstr\"om case, i.e., the absence of
{\it fine-tunnings} for the intrinsic tension.

In spite of the existence of {\it fine-tunnings}, the fact is that shortcuts
appear and consequences are manifold.
As mentioned in \cite{ishihara,cald-lang}, the existence of
shortcuts could partially solve the horizon problem.
We should notice that our set of conditions to obtain shortcuts in
AdS-Schwarzschild and 
AdS-Reissner-Nordstr\"om bulks with protected singularities impose a
restriction on the size of the universe, namely $z_0 \sim l$, which
corresponds to a primeval universe. So the results shown in the
present paper could contribute to the solution of this important
problem.
  
We can also point out experimental consequences. In particular,
gravitational waves advanced with respect to photons might be found in
the proposed gravitational antennas under way, in case we find a model
for the Universe with a physical size.

\bigskip

{\bf Acknowledgements:} This work has been supported by Funda\c c\~ao
de Amparo \`a Pesquisa do Estado de S\~ao Paulo {\bf (FAPESP)} and Conselho
Nacional de Desenvolvimento Cient\'\i fico e Tecnol\'ogico {\bf
(CNPq)}, Brazil.

\end{document}